\documentclass[aps,prd,floats,floatfix,preprintnumbers,superscriptaddress,nofootinbib]{revtex4-2}

\usepackage{graphicx} 
\usepackage{epstopdf} 
\usepackage[normalem]{ulem}
\usepackage{amssymb}
\usepackage{amsmath}
\usepackage[hidelinks]{hyperref}
\usepackage{color}
\usepackage{xcolor,colortbl}
\usepackage[utf8]{inputenc}
\usepackage{slashed}

\newcommand{\be}{\begin{equation}}
\newcommand{\ee}{\end{equation}}

\newcommand{\rmd}{\mathrm{d}}
\newcommand{\rme}{\mathrm{e}}
\newcommand{\rmi}{\mathrm{i}}
\newcommand{\calA}{\mathcal{A}}

\newcommand{\calP}{\mathcal{P}}

\newcommand{\calM}{\mathcal{M}}

\newcommand{\calN}{\mathcal{N}}

\newcommand{\calO}{\mathcal{O}}

\newcommand{\calU}{\mathcal{U}}

\newcommand{\konp}{\boldsymbol{k}_{1\perp}}
\newcommand{\ktwp}{\boldsymbol{k}_{2\perp}}
\newcommand{\kthp}{\boldsymbol{k}_{3\perp}}

\newcommand{\rp}{\boldsymbol{r}_{\perp}}

\newcommand{\bp}{\boldsymbol{b}_{\perp}}

\newcommand{\lp}{\boldsymbol{l}_{\perp}}

\newcommand{\up}{\boldsymbol{u}_{\perp}}
\newcommand{\vp}{\boldsymbol{v}_{\perp}}
\newcommand{\xp}{\boldsymbol{x}_{\perp}}
\newcommand{\yp}{\boldsymbol{y}_{\perp}}

\newcommand{\Sp}{\boldsymbol{S}_{\perp}}

\newcommand{\kp}{\boldsymbol{k}_{\perp}}

\newcommand{\delp}{\boldsymbol{\Delta}_{\perp}}

\newcommand{\epsp}{\boldsymbol{\epsilon}_{\perp}}

\begin{document}
\preprint{ZTF-EP-24-06}

\title{Gluon Sivers function from forward exclusive $\chi_{c1}$ photoproduction on unpolarized protons}

\author{Sanjin Beni\' c}
\affiliation{Department of Physics, Faculty of Science, University of Zagreb, Bijenička c. 32, 10000 Zagreb, Croatia}

\author{Adrian Dumitru}
\affiliation{Department of Natural Sciences, Baruch College, CUNY,
17 Lexington Avenue, New York, NY 10010, USA}
\affiliation{The Graduate School and University Center, The City University of New York, 365 Fifth Avenue, New York, NY 10016, USA}

\author{Leszek Motyka}
\affiliation{Jagiellonian University, Institute of Theoretical Physics,
Łojasiewicza 11, 30-348 Kraków, Poland}

\author{Tomasz Stebel}
\affiliation{Jagiellonian University, Institute of Theoretical Physics,
Łojasiewicza 11, 30-348 Kraków, Poland}

\begin{abstract}
Exclusive production of a $\chi_{c1}$ axial vector quarkonia in photon-proton
scattering at high energies requires a $C$-odd $t$-channel
exchange. In the limit of vanishing momentum transfer this occurs
either via the exchange of a photon, the Primakoff process, where the
spin of the proton does not change.  For axial-vector meson
production, as a consequence of the Landau-Yang theorem, the Primakoff
cross section is finite as $t \to 0$.  Alternatively, a $C$-odd spin
dependent Odderon can be exchanged, which involves a spin flip of the
proton. The resulting cross section is related to the square of the collinear trigluon correlator or the
$k_\perp$-moment of the gluon Sivers function. Using two models for the gluon Sivers function
from the literature we compute the ratio of Sivers to Primakoff cross
sections and the angular coefficient $\lambda_\theta$ governing the angular distribution of the $\chi_{c1}
\to J/\psi + \gamma$ decay as functions of $x$. We point out that
these observables constrain the magnitude of
the gluon Sivers function at small $x$ which could be accessed in electron-proton
scattering and ultraperipheral proton-proton and nucleus-proton collisions.
\end{abstract}

\maketitle

\section{Introduction} 
In  our previous work \cite{Benic:2024pqe} we have computed the exclusive cross section of axial vector $\chi_{c1}$ quarkonia: $\gamma^*(q) p(P) \to \chi_{c1}(\Delta) p(P')$. This arises from tri-gluon (or, Odderon) exchange with the proton, interfering with the photon exchange (the Primakoff process). The focus was on the so-called spin-independent Odderon \cite{Boussarie:2019vmk} which is important at finite momentum transfer $t = (P - P')^2 \approx -\delp^2$.
For $t \to 0$, on the other hand, the corresponding cross section vanishes due to odd parity of the spin independent Odderon amplitude under $\delp \to -\delp$. 

In near-forward kinematics the Primakoff process becomes the major background to Odderon searches. Nevertheless, the $\chi_{c1}$ is special because, thanks to the Landau-Yang theorem~\cite{Landau:1948kw,Yang:1950rg}, the near-forward Coulomb tail ($\sim 1/|t|$) of the Primakoff cross section, in the photoproduction $q^2 = - Q^2 \to 0$ 
limit\footnote{We require here the scale hierarchy $Q^2\ll |t| \lesssim 1/R^2_p$ where
$R_p\simeq 1$~fm is the radius of the proton.}, is screened \cite{Benic:2024pqe}, and the cross section
remains finite. In the non-relativistic (NR) $m_c\to\infty$ limit we simply have~\cite{Benic:2024pqe,Jia:2022oyl}
\be
\lim_{t \to 0}\frac{\rmd \sigma_{\rm Prim}}{\rmd |t|} = \frac{3\pi q_c^4 \alpha^3 N_c |R'(0)|^2 |F_1(0)|^2}{m_c^9}\,.
\label{eq:primch}
\ee
Here $R'(0)$ is the derivative of the NR wave function at the origin ($\chi_{c1}$ is classified as $P$-wave quarkonium \cite{ParticleDataGroup:2022pth}), $m_c$ is the charm quark mass, $\alpha$ is the electromagnetic fine structure constant, with $q_c = 2/3$ the quark fractional charge, and $N_c = 3$ denotes the number of colors. The Dirac form-factor $F_1(t)$ is evaluated at $t = 0$. Note that this result is also independent of the collision energy $W^2 = (q + P)^2$ in the eikonal limit, $W^2 \gg |t|$.

It has been pointed out in \cite{Boussarie:2019vmk} (see also \cite{Ma:2003py}) that the near-forward region of the {\it unpolarized} $\gamma^* p$ cross section provides an opportunity to constrain the spin-dependent Odderon~\cite{Boussarie:2019vmk,Boer:2015pni}, that is, the gluon Sivers function. The gluon Sivers function represents an azimuthally asymmetric transverse momentum distribution (TMD) of gluons in the proton that is expected in collisions off a transversely polarized proton \cite{Burkardt:2004ur,Meissner:2007rx,Boer:2015vso}. This function
represents a key element in our understanding of the structure of the proton; to date, it is rather poorly constrained,
as will be clear from fig.~\ref{fig:primsiv2} below.
The determination of the gluon Sivers function at high energies
represents one of the key ``golden measurements" to be performed at the
upcoming electron-ion collider (EIC)~\cite{Proceedings:2020eah,AbdulKhalek:2021gbh,Achenbach:2023pba}.

\begin{figure}[htb]
  \begin{center}
  \includegraphics[scale = 0.5]{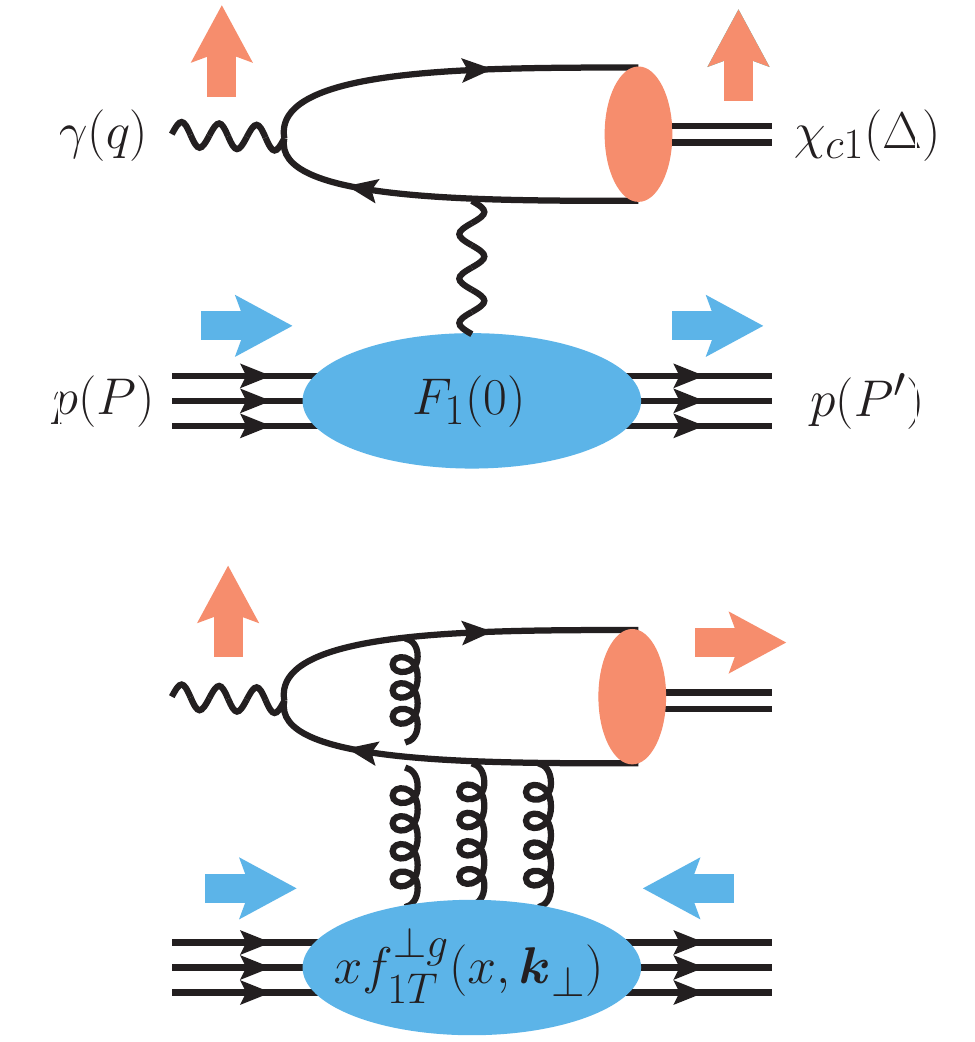}
  \end{center}
  \caption{Representative diagrams illustrating the Primakoff (top) and the Sivers (bottom) contributions to the exclusive $\gamma(q) p(P) \to \chi_c(\Delta) p(P')$ process. The arrows indicate the helicity transfer in the respective channels.}
  \label{fig:diags}
\end{figure}

The traditional strategy for measuring the gluon Sivers function is through single spin asymmetries (SSAs) in single-inclusive $e p^\uparrow$ \cite{Kang:2008qh,Beppu:2010qn,Zhou:2013gsa,Godbole:2017fab,Zheng:2018ssm,Chen:2023hvu} or $p p^\uparrow$ collisions \cite{Anselmino:2004nk,DAlesio:2015fwo,Kato:2024vzt} with heavy quark production. Considering instead exclusive $C$-even meson production, a finite contribution arises to the {\it unpolarized} $e p$ cross section from the gluon Sivers function  when $t \to 0$ \cite{Boussarie:2019vmk}. The simple reason is the possibility of a proton helicity flip as associated with the Sivers function; this is illustrated by the lower diagram in Fig~\ref{fig:diags}. We argue that measuring axial quarkonia such as $\chi_{c1}$ for the same near-forward kinematics has a unique advantage over other mesons ($\pi^0$ or $\eta$, but also $a_2$, $f_2$,\dots) or even other quarkonia ($\eta_c$, $\chi_{c0}$, $\chi_{c2}$,\dots). In these cases the Primakoff cross section blows up at $t\to 0$, and it will {\it always} dominate the Sivers contribution. By contrast, for axial quarkonia such as $\chi_{c1}$, the Primakoff process gives a finite contribution \eqref{eq:primch}. Depending on the relative magnitudes of the Sivers vs.\ Primakoff channels there is potentially a chance to become directly sensitive to the gluon Sivers function, without the need of subtracting the Primakoff background at all.

We propose another promising opportunity to constrain the gluon Sivers function 
via the polarization of $\chi_{c1}$ in the Sivers and the Primakoff production channels, respectively. 
The argument is as follows. Since the photon exchange at high energies cannot flip the proton helicity (at $t \to 0$ and $Q^2 \to 0$), conservation of helicity dictates that the $\chi_{c1}$ from the Primakoff channel must have {\it transverse} polarization. By contrast, the spin dependent Odderon flips the proton helicity and so the produced $\chi_{c1}$ must have {\it longitudinal} polarization. This way, the polarization of $\chi_{c1}$ is directly sensitive to the gluon Sivers function. The different polarization states can be accessed through the angular distribution of the $\chi_{c1}$ decay products for which we provide an example below. The helicity transfer from the initial photon to the final $\chi_{c1}$ is illustrated for both channels in fig.~\ref{fig:diags}, with red arrows representing photon or $\chi_{c1}$ polarization vectors, and the blue arrows stand for proton helicity.

\section{Exclusive $\gamma p \to \chi_{c1} p$ cross section} 


The amplitude for the process $\gamma p \to \chi_{c1} p$ is given by~\cite{Benic:2024pqe}
\be
\begin{split}
\langle \calM_{\lambda_p\lambda_\gamma,\lambda'_p\lambda_\chi}\rangle & = 2 q^- N_c\int_{\konp\ktwp} (2\pi)^2\delta^{(2)}(\konp + \ktwp - \delp)\\
&\times \int_{\xp\yp} \rme^{-\rmi \konp \cdot\xp - \rmi \ktwp\cdot\yp}\rmi \calO_{\lambda_p \lambda_p'}(\xp,\yp)\calA_{\lambda_\gamma\lambda_\chi}(\konp,\ktwp)\,,
\end{split}
\label{eq:ampchi}
\ee
where $\calO_{\lambda_p \lambda_p'}(\xp,\yp)$ is the helicity dependent Odderon exchange amplitude 
\be
\calO_{\lambda_p \lambda_p'}(\xp,\yp) = \frac{1}{N_c} {\rm Im}\left[{\rm tr}\langle V^\dag(\xp)V(\yp)\rangle\right] \,.
\label{eq:Odef}
\ee
Here $\langle\dots \rangle = \langle P'\lambda_p' |\dots|P \lambda_p\rangle/\langle P \lambda_p|P \lambda_p\rangle$
denotes a matrix element in the proton state, while $\xp = \bp + \frac{\rp}{2}$, $\yp = \bp - \frac{\rp}{2}$ are transverse coordinates, and $V(\xp) \equiv \calP \exp\left[-\rmi g \int_{-\infty}^\infty \rmd x^- A^+(x^-,\xp)\right]$ are Wilson lines describing scattering of the $c$ quarks off the gluon fields $A^+(x^-,\xp) = A_a^{+}(x^-,\xp) t_a$ in the proton in the eikonal (high-energy) limit. $\lambda_p$ ($\lambda_p'$) are the helicities of the incoming (outgoing) proton. The Wilson lines carry momenta $\konp$ and $\ktwp$ out of the proton and into the $\gamma-\chi_{c_1}$ impact factor $\calA_{\lambda_\gamma\lambda_\chi}(\konp,\ktwp)$,
\be
\calA_{\lambda_\gamma\lambda_\chi}(\konp,\ktwp) \equiv e q_c \int_0^1 \frac{\rmd z}{4\pi}\int_{\lp} \frac{\phi_\chi(\lp,z)}{z\bar{z}}\frac{A_{\lambda_\gamma\lambda_\chi}(\lp - \bar{z}\konp + z\ktwp,\lp,z)}{(\lp - \bar{z}\konp + z\ktwp)^2 + m_c^2 + z\bar{z}Q^2}\,,
\ee
where $\lambda_\gamma$ is the helicity of the incoming photon, $\phi_\chi(\lp,z)$ is the  $\chi_{c1}$ light-cone wave function, and $\lambda_\chi$ its helicity. An explicit expression for the function $A_{\lambda_\gamma\lambda_\chi}(\lp - \bar{z}\konp + z\ktwp,\lp,z)$ has been provided in eq.~(B1) of ref.~\cite{Benic:2024pqe}. We use the notation $\int_{\xp} = \int \rmd^2 \xp$ and $\int_{\kp} = \int\frac{\rmd^2 \kp}{(2\pi)^2}$ for transverse coordinate and momentum integrals, respectively.

We now demonstrate explicitly that the amplitude is proportional to the gluon Sivers function (more precisely to its first moment). For this purpose we start by expanding the Wilson lines in \eqref{eq:Odef} to cubic order in the static
mode of $A^+$ which leads to the well-known expression \cite{Hatta:2005as,Dumitru:2019qec}
\be
\begin{split}
\calO_{\lambda_p\lambda_p'}(\xp,\yp) & = - \frac{g^3}{24 N_c} d_{abc}\big\langle 3 \alpha_a(\xp)\alpha_b(\yp)\alpha_c(\yp) - 3 \alpha_a(\xp)\alpha_b(\xp)\alpha_c(\yp)\\
& + \alpha_a(\xp)\alpha_b(\xp)\alpha_c(\xp) - \alpha_a(\yp)\alpha_b(\yp)\alpha_c(\yp)\big\rangle\,,
\end{split}
\label{eq:oddexp}
\ee
where 
\be
\alpha_a(\xp) \equiv \int_{-\infty}^\infty \rmd x^- A_a^+(x^-,\xp)\,.
\label{eq:alpa}
\ee 
Inserting \eqref{eq:oddexp} into \eqref{eq:ampchi} we get
\be
\begin{split}
\langle &\calM_{\lambda_p\lambda_\gamma,\lambda_p'\lambda_\chi}\rangle = - 2q^- \frac{\rmi g^3}{24} d_{abc} \int_{\konp\ktwp\kthp} (2\pi)^2\delta^{(2)}(\konp + \ktwp + \kthp - \delp) \left\langle\alpha_a(\konp)\alpha_b(\ktwp)\alpha_c(\kthp)\right\rangle\\
&\times \Big[\calA_{\lambda_\gamma\lambda_\chi}(\konp,\ktwp+\kthp) - \calA_{\lambda_\gamma\lambda_\chi}(\ktwp+\kthp,\konp) + \calA_{\lambda_\gamma\lambda_\chi}(\ktwp,\konp+\kthp) - \calA_{\lambda_\gamma\lambda_\chi}(\konp+\kthp,\ktwp)\\
& + \calA_{\lambda_\gamma\lambda_\chi}(\kthp,\konp+\ktwp) - \calA_{\lambda_\gamma\lambda_\chi}(\konp+\ktwp,\kthp) + \calA_{\lambda_\gamma\lambda_\chi}(\konp+\ktwp+\kthp,0) - \calA_{\lambda_\gamma\lambda_\chi}(0,\konp+\ktwp+\kthp)\Big]\,,
\end{split}
\label{eq:impactodd}
\ee
with the combination of impact factors in the square bracket in agreement with ref.~\cite{Bartels:2007aa}. In \eqref{eq:impactodd}, $\alpha_a(\kp)$ is the Fourier transform of $\alpha_a(\xp)$.

We now simplify the above exact expression eq.~(\ref{eq:impactodd}) in several steps. First, we take the photoproduction, $Q^2 \to 0$, as well as the forward, $\delp \to 0$, limits. As a consequence of helicity conservation only the amplitude for longitudinally polarized $\chi_{c1}$ survives, $\lambda_\chi = 0$. We also employ the collinear limit on the proton side ($\kp^2 \ll m_c^2$) and the NR limit ($\lp^2 \ll m_c^2$) to reveal the underlying collinear proton distribution. We then find
\be
\calA_{\lambda_\gamma 0}(\konp,\ktwp) \approx \frac{\sqrt{3}}{16\sqrt{\pi}}\frac{e q_c R'(0)}{m_c^{11/2}\sqrt{N_c}} \epsp^{\lambda_\gamma}\times(\ktwp - \konp) (\ktwp - \konp)^2 \,.
\label{eq:impac}
\ee
Here we used the relation \cite{Benic:2024pqe}
\be
\int_0^1\frac{\rmd z}{4\pi}\int_{\lp} \lp^2\frac{\phi_\chi(\lp,z)}{z\bar{z}} = \frac{\sqrt{3}}{4\sqrt{\pi}}\frac{R'(0)}{\sqrt{m_c} \sqrt{N_c}}
\ee
to trade the transverse and longitudinal momentum integrals over the light-cone wave function for the derivative at the
origin of the NR radial wave function of $\chi_{c1}$. Furthermore,
 $\epsp^\lambda = (-\lambda,-\rmi)/\sqrt{2}$, and the 2D cross product is defined as $\up\times \vp = \epsilon^{ij}u_\perp^i v_\perp^j = u_\perp^1 v_\perp^2 - u_\perp^2 v_\perp^1$.
Inserting \eqref{eq:impac} into \eqref{eq:impactodd} leads to 
\be
\begin{split}
\langle\calM_{\lambda_p\lambda_\gamma,\lambda_p' 0}\rangle & = q^- \frac{\rmi \sqrt{3}}{4\sqrt{\pi}} g^3 d_{abc} \frac{e q_c R'(0)}{m_c^{11/2}\sqrt{N_c}}\epsilon^{ij}\delta^{mn} \int_{\konp\ktwp\kthp} (2\pi)^2\delta^{(2)}(\konp + \ktwp + \kthp)\\
&\times\epsilon_\perp^{\lambda_\gamma,i} k_{1\perp}^j k_{2\perp}^m k_{3\perp}^n \langle\alpha_a(\konp)\alpha_b(\ktwp)\alpha_c(\kthp)\rangle\,.
\end{split}
\label{eq:amphq}
\ee
The individual gluon momenta can now be employed to form the field strength tensor, $F_a^{i+} \approx \partial^i A_a^+$, via partial integrations,  which will turn the soft matrix element in \eqref{eq:amphq} into a tri-gluon correlator. The $d$-type tri-gluon correlator is commonly defined~\cite{Ji:1992eu,Beppu:2010qn,Koike:2019zxc}  for a {\it transversely} polarized proton with spin $\Sp$ as (our notation follows ref.~\cite{Beppu:2010qn})
\be
\begin{split}
\frac{g}{P^+} &\int\frac{\rmd y^-}{2\pi}\int\frac{\rmd z^-}{2\pi}\rme^{\rmi x_1 P^+ z^-}\rme^{\rmi (x_2 - x_1) P^+ y^-}\langle P \Sp|d_{bca}F_b^{j +}(0) F_c^{k +}(y^-)F_a^{i +}(z^-) |P \Sp\rangle\\
& = - 2 M_p  S_\perp^l \left[O(x_1,x_2)\delta^{ij}\epsilon^{kl} + O(x_2,x_2 - x_1)\delta^{jk}\epsilon^{il} + O(x_1,x_1 - x_2)\delta^{ik}\epsilon^{jl} \right]\,,
\end{split}
\label{eq:trig}
\ee
where $M_p$ the proton mass. Comparing to e.~g. eqs.~(1) and (7) in \cite{Beppu:2010qn} we also have the convention $\epsilon^{ij} = -\epsilon^{+-ij}$ with $\epsilon^{+-12} = -1 = - \epsilon^{0123}$. The transverse spin and the helicity basis are related by~\cite{Meissner:2007rx}
\be
|P\Sp \rangle = \frac{1}{\sqrt{2}}(|P, \lambda_p = +1\rangle + \rme^{\rmi \phi_S} |P, \lambda_p = -1\rangle)\,,
\label{eq:basis}
\ee
with $\phi_S$ the angle of $\Sp$. Using \eqref{eq:basis} in \eqref{eq:trig} amounts to a replacement $\Sp \to - \sqrt{2}\lambda_p\epsp^{\lambda_p} \delta_{\lambda_p,-\lambda_p'}$ in terms of the helicity basis. We can now parametrize the amplitude as
\be
\langle\calM_{\lambda_p\lambda_\gamma,\lambda_p'\lambda_\chi}\rangle = q^- \lambda_\gamma \delta_{\lambda_\gamma,-\lambda_p}\delta_{\lambda_p,-\lambda_p'}\calM_{\rm Siv}\,.
\label{eq:Mparam}
\ee
Inserting \eqref{eq:trig} into 
\eqref{eq:amphq} for small $x \equiv x_1 \approx x_2 \approx 0$ yields
\be
\calM_{\rm Siv} = 2\sqrt{6} \pi^{3/2} e q_c \alpha_S \frac{1}{\sqrt{N_c}}\frac{R'(0) M_p}{m_c^{11/2}} 8\pi O(x)\,,
\ee
where we write $O(x) \equiv O(x,0)\approx O(x,x)$. 

On the other hand, the dipole-type gluon Sivers function $x f_{1T}^{\perp g}(x,\kp)$ is defined through the following correlator \cite{Mulders:2000sh,Boer:2016xqr}
\be
\begin{split}
\frac{2}{P^+}\int \frac{\rmd v^- \rmd^2 \vp}{(2\pi)^3} \rme^{\rmi x P^+ v^-} \rme^{-\rmi \kp\cdot \vp} & \langle P \Sp| {\rm tr}\big[F^{i+}(-v/2) \calU^{[+]}_{[-\frac{v}{2},\frac{v}{2}]} F^{i+}(v/2)\calU^{[-]\dagger}_{[-\frac{v}{2},\frac{v}{2}]}\big]|P \Sp\rangle\\
& = x f_1^g(x,\kp) - \frac{(\kp\times \Sp)}{M_p} x f_{1T}^{\perp g}(x,\kp) \,,
\label{eq:gltsiv}
\end{split}
\ee 
where $\calU^{[\pm]}_{[-\frac{v}{2},\frac{v}{2}]}$ are staple-shaped gauge links required to make the operator gauge invariant (see App.~\ref{sec:appa} for the definition) and $f_1^g(x,\kp)$ is the unpolarized TMD. Defining the first $k_\perp$-moment
\be
x f_{1T}^{\perp(1) g}(x,\mu^2) \equiv \int^{\mu^2} \rmd^2 \kp \frac{\kp^2}{2 M_p^2} x f_{1T}^{\perp g}(x,\kp)\,,
\label{eq:momsiv1}
\ee
with $\mu^2$ being the factorization scale, the connection to $O(x)$ from \eqref{eq:trig} at small-$x$ is established
through~\cite{Zhou:2013gsa,Boer:2015pni}
\be
x f_{1T}^{\perp (1)g}(x) = 8\pi O(x)\,.
\label{eq:sivmomO}
\ee
In App.~\ref{sec:appa} we derive \eqref{eq:sivmomO} by using the connection between the TMD and the dipole operator ${\rm tr}(V^\dag(\xp)V(\yp))/N_c$ at small $x$~\cite{Dominguez:2011wm}, and we furthermore clarify that $O(x)$ also governs the small-$x$ behavior of two more gluon TMDs of a transversely polarized proton: $h_{1T}^g(x,\kp)$ and $h_{1T}^{\perp g}(x,\kp)$ \cite{Boer:2015pni}.  Using \eqref{eq:Mparam}, the cross section in the NR limit is
\be
\lim_{t\to 0} \frac{\rmd \sigma_{\rm Siv}}{\rmd |t|} = \frac{1}{32\pi}|\calM_{\rm Siv}|^2 = \frac{3\pi^3 q_c^2 \alpha\alpha_S^2 M_p^2 \, |R'(0)|^2 \, |x f_{1T}^{\perp (1) g}(x,\mu^2)|^2}{N_c m_c^{11}}\,.
\label{eq:sivch}
\ee
Eq.~\eqref{eq:sivch} parallels the classic result for the exclusive $J/\psi$ production cross section at $t \to 0$ which is proportional to the square of the unpolarized gluon distribution function \cite{Ryskin:1995hz}. Comparing to the Primakoff cross section in \eqref{eq:primch}, eq.~\eqref{eq:sivch} is suppressed by two additional powers of $m_c$ due to the twist-3 nature of the Sivers function.

\begin{figure}[htb]
  \begin{center}
  \includegraphics[scale = 0.8]{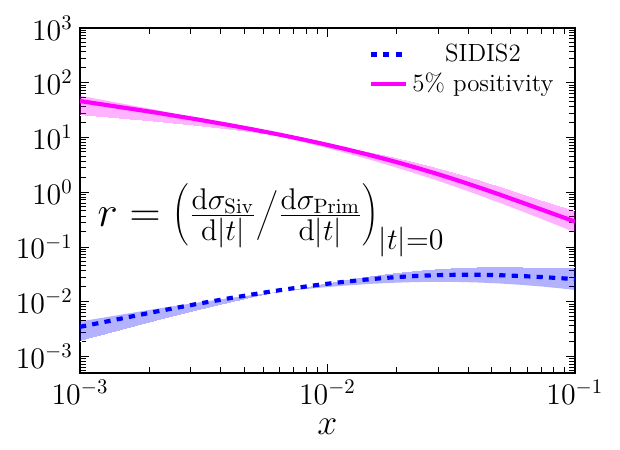}
  \end{center}
  \caption{Ratio of the $\gamma p \to \chi_{c1} p$ cross sections in the Sivers to Primakoff channels, in the $t \to 0$ limit, as a function of diffractive $x$. These processes involve $C$-odd tri-gluon (Odderon)
  and photon exchanges, respectively.
The bands correspond to variation of the factorization scale $\mu$ by a factor of 2 around the central value $\mu = 2 m_c$.}
  \label{fig:primsiv2}
\end{figure}
In practice only the sum of the Sivers and the Primakoff cross sections can be measured. Nevertheless, we argue that the two contributions can be disentangled as follows. Firstly, thanks to the spin flip nature of the Sivers function there is no interference of the amplitudes. Secondly, at high energy only the gluon Sivers function is expected to exhibit $x$ dependence; recall that \eqref{eq:primch} is completely independent of $x$. 
Whether the $x$ dependence in the total cross section would be significant or not depends, of course, on the relative magnitudes of the Primakoff and Sivers components. It is useful to consider the ratio
\be
r \equiv \left(\frac{\rmd \sigma_{\rm Siv}}{\rmd |t|} \Big/\frac{\rmd \sigma_{\rm Prim}}{\rmd |t|}\right)_{t = 0}\,,
\label{eq:r}
\ee
where the Sivers cross section is in general given by eq.~\eqref{eq:ampchi}, and the Primakoff cross section follows from eqs.~(33) and (43) in \cite{Benic:2024pqe}. The NR limit is again very illustrative:
\be
r \to \frac{4\pi^2}{q_c^2 N_c^2} \frac{\alpha_S^2}{\alpha^2} \frac{M^2_p}{M^2_{\chi}}
|x f_{1T}^{\perp (1) g}(x,\mu^2)|^2\,,
 \label{eq:NR-r}
\ee
with $R'(0)$ cancelling in the ratio! In general, we can expect the uncertainty in modelling the non-perturbative $\chi_{c1}$ wave function to partially cancel in $r$. Here $M_\chi = 2 m_c$ is the mass of $\chi_{c1}$ in the NR limit. 

In fig.~\ref{fig:primsiv2} we show numerical results for \eqref{eq:NR-r} using
two models for $x f_{1T}^{\perp g}(x,\kp)$. 
The model labelled ``SIDIS2" and shown with the dashed blue curve was obtained from a fit to transverse single spin asymmetries
measured at RHIC~\cite{DAlesio:2015fwo}. Its first Sivers moment is \cite{DAlesio:2015fwo}
\be
|f_{1T}^{\perp (1) g}(x,\mu^2)| = \sqrt{\frac{\rme}{2}} \rho^2 \sqrt{\frac{1-\rho}{\rho}}\frac{\sqrt{\langle k_\perp^2\rangle}}{M_p} \calN_g(x) g(x,\mu^2)\,,
\label{eq:sidis2}
\ee
where $\calN_g(x) = N_g x^\alpha (1-x)^\beta (\alpha + \beta)^{\alpha + \beta}/(\alpha^\alpha \beta^\beta)$ with $N_g = 0.05$, $\alpha = 0.8$, $\beta = 1.4$, $\rho = 0.576$, $\langle k_\perp^2\rangle = 0.25$ GeV$^2$, and $g(x)$ is the unpolarized gluon PDF. The label ``5\% positivity" (full magenta curve) represents a model~\cite{Zheng:2018ssm} where the magnitude of the Sivers function is set at
$5\%$ of the function that saturates the positivity bound 
within the model of \cite{Anselmino:2004nk}:
\be
\begin{split}
|f_{1T}^{\perp (1) g}(x,\mu^2)| & = 0.05\times \frac{\sigma}{M_p}\left[1 + \rho\rme^{\rho} {\rm Ei}(-\rho)\right] g(x,\mu^2)\,,
\label{eq:5posit}
\end{split}
\ee
 where ${\rm Ei}(\rho)$ is the exponential integral, with $\rho = \sigma^2/\langle k_\perp^2\rangle$. In the numerical computation we employ $\sigma^2 = \langle k_\perp^2\rangle$ and $\sigma = 0.8$ GeV \cite{Anselmino:2004nk}.

The difference between the two models in fig.~\ref{fig:primsiv2} illustrates the large uncertainty in our current knowledge of $f_{1T}^{\perp g}(x,\kp)$. The bands were obtained by varying the factorization scale $\mu$ by a factor of 2 around the central value $\mu = 2 m_c$. Clearly, neither curve in fig.~\ref{fig:primsiv2} represents a reliable
theoretical prediction; rather, their purpose is to outline opposing scenarios of large and small Sivers contribution, as
well as strong vs.\ weak dependence of $r$ on $x$.
The 5\% positivity model indicates a potentially large contribution from the Sivers component to the $\chi_{c1}$ cross section, emphasizing that a measurement of the exclusive $\gamma p \to \chi_{c1} p$ cross section, in particular of its $x$ dependence, can be used to constrain the gluon Sivers function.  A computation based on eq.~\eqref{eq:ampchi} gives a factor of 2.0 (3.0) lower Sivers cross section for the SIDIS2 (5\% positivity) models. For the Primakoff cross section the analogous ratio is 1.12, in line with expectations from relativistic corrections \cite{Lappi:2020ufv}. The larger difference in case of the Sivers cross section originates from finite-$k_\perp$ corrections on top of relativistic corrections, both of which are included in \eqref{eq:ampchi}.


It is interesting to contrast the above two phenomenological models to a small-$x$ model for the spin dependent Odderon \cite{Zhou:2013gsa,Lappi:2016gqe,Yao:2018vcg} which follows from a group theory constraint. 
The trace properties of the special unitary matrix in \eqref{eq:Odef} restrict the magnitude of the Odderon to 
$|\calO(\rp)| \le \calN(\rp)^{3/2}/3$, in the weak field limit where the $C$-even (Pomeron) amplitude $\calN(\rp)$ is small.
The form and magnitude of $\calN(\rp)$ 
at the initial condition $x_0 \simeq 0.01$ of the small-$x$ dipole model \cite{Kovchegov:2003dm,Jeon:2005cf,Hatta:2005as,Zhou:2013gsa}
are fairly well 
constrained~\cite{Kowalski:2006hc,Albacete:2009fh,Lappi:2013zma,Dumitru:2023sjd,Casuga:2023dcf}.
Relating $\calN(\rp)$ (in the double-log regime)
to the leading-twist collinear gluon density $xg(x)$~\cite{Bartels:2002cj,Kowalski:2006hc}
we obtain that this ``maximal Odderon" corresponds to the Sivers moment
\be
|f_{1T}^{\perp (1) g}(x) | = \frac{1}{3} \frac{Q_S(x)}{M_p} g(x)~.
\label{eq:f1Tsmallx}
\ee
$Q_S(x)$ denotes the
intrinsic transverse momentum scale of the proton at $x$.
With $Q_S(x_0)=0.25 - 0.5$~GeV, the upper limit for $|f_{1T}^{\perp (1) g}(x_0)|$ 
reaches $9-18\%$ of the unpolarized gluon density. Comparing to \eqref{eq:5posit}, but at 100\% of the positivity bound (at the TMD level), $f^{\perp (1) g}_{1T}(x)$ is about 35\% of $g(x)$. Interestingly, the group theory bound, 
valid at small-$x$, is more stringent than the positivity bound. 

The above models do not represent solutions of the small-$x$ evolution equations. Rather, they should be understood as models of the initial condition\footnote{Alternatively, one may
compute the gluon Sivers function at moderately small $x$ from a model light-cone wave function for the
proton, as has been done for the impact parameter dependent eikonal Odderon~\cite{Dumitru:2018vpr,Dumitru:2019qec,Dumitru:2021tqp}. This is work in progress which will be reported elsewhere.}. Computing the Sivers function at very small $x$, say $x \sim 10^{-3}$ or below, requires
resummation of perturbative corrections involving additional powers of $\alpha_S\log 1/x$.
In contrast to the Pomeron, the small-$x$ linear evolution of the Odderon \cite{Bartels:1999yt} leads to a roughly constant energy dependence, $\sim 1/x^0$, and suppressed by non-linear unitarity corrections \cite{Kovchegov:2003dm,Motyka:2005ep,Hatta:2005as,Lappi:2016gqe,Yao:2018vcg,Benic:2023ybl}.

At this point it is useful to provide an estimate of the magnitude of the Primakoff cross section for the EIC but also for the ultra-peripheral collisions (UPCs) at RHIC and the LHC. Taking $|R'(0)|^2 = 0.1296$ GeV$^5$ \cite{Eichten:2019hbb} 
we obtain from \eqref{eq:primch} that 
$(\rmd \sigma_{\rm Prim}/\rmd |t|)_{t = 0} \approx 0.69$ pb/GeV$^2$. For $pA$ UPCs we convolute the $\gamma-p$ cross section with the photon flux factor \cite{Baltz:2007kq} and obtain $0.05$ nb/GeV$^2$ ($0.79$ nb/GeV$^2$) at $\sqrt{s} = 200$ GeV ($5.02$ TeV) RHIC (LHC) energy.


\begin{figure}[htb]
  \begin{center}
  \includegraphics[scale = 0.7]{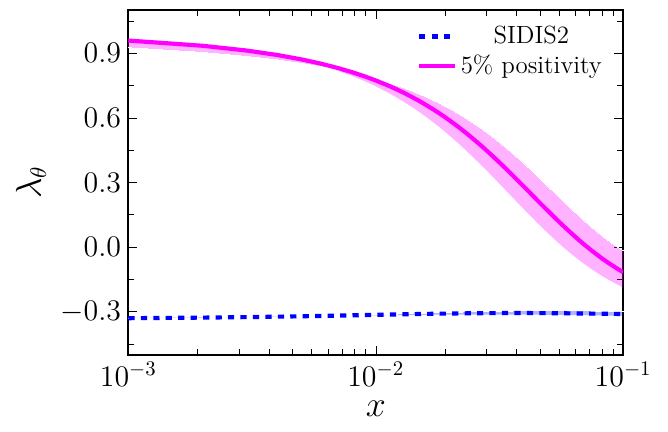}
  \end{center}
  \caption{Angular coefficient $\lambda_\theta$ as a function of $x$ based on different models of the gluon Sivers function (see text). The bands correspond to variation of the factorization scale $\mu$ by a factor of 2 around the central value $\mu = 2 m_c$.}
  \label{fig:lamth}
\end{figure}

\section{Angular distributions in $\chi_{c1} \to J/\psi\gamma$ decay}
As already mentioned, the Primakoff process produces transversely polarized $\chi_{c1}$'s
while those from the Sivers channel are longitudinally polarized. In order to distinguish the polarization we focus on the dominant decay channel, $\chi_{c1} \to J/\psi + \gamma$, assuming a pure electric dipole (E1) transition, for simplicity.
The angular distribution of $J/\psi$'s is~\cite{Faccioli:2010ji,Faccioli:2011be,Shao:2012fs} 
\be
W(\theta,\phi) \propto \frac{N}{3 + \lambda_\theta} \big(1 + \lambda_\theta\cos^2\theta + \dots\big)\,.
\label{eq:angW}
\ee
The ellipses in \eqref{eq:angW} correspond to four additional angular coefficients $\lambda_i$ that 
drop out by explicit calculation.
The polarization angles $\theta$ and $\phi$ are frame-dependent; we assume the Collins-Soper frame. 
Since our calculation \cite{Benic:2024pqe} has been performed in the $\gamma p$ center-of-mass frame, with the helicity quantized along the $z$-direction, we have confirmed that the general choice of the Collins-Soper polarization vectors \cite{Beneke:1998re} collapses to that in \cite{Benic:2024pqe} when $t \to 0$. 
The simple reason is that the Collins-Soper frame and the $\gamma p$ center-of-mass frame are related by a longitudinal boost when $t \to 0$. 
All the angular coefficients $\lambda_i$ can be expressed in terms of the spin density matrix
$\rho_{\lambda'_\chi,\lambda_\chi} \propto 
\sum_{\lambda_\gamma\lambda_p\lambda'_p}\calM_{\lambda_p\lambda_\gamma,\lambda_p'\lambda'_\chi}^*\calM_{\lambda_p\lambda_\gamma,\lambda_p'\lambda_\chi}$. 
The $\lambda_\theta$ coefficient is given by~\cite{Faccioli:2010ji,Faccioli:2011be,Shao:2012fs}
\be
\lambda_\theta = \frac{3\rho_{0,0} - N}{3 N - \rho_{0,0}}\,,
\label{eq:lamth}
\ee
where $N = \rho_{1,1} +\rho_{0,0} + \rho_{-1,-1}$. 
The remaining angular coefficients are all proportional to off-diagional entries of the density matrix (unlike $\lambda_\theta$) which are zero in both the Sivers and the Primakoff channels. The diagonal entries yield
\be
\begin{split}
&\rho_{0,0} \propto 16\pi \lim_{|t| \to 0} \frac{\rmd\sigma_{\rm Siv}}{\rmd |t|}\,,\\
&\rho_{1,1} = \rho_{-1,-1} \propto 8\pi \lim_{|t| \to 0} \frac{\rmd\sigma_{\rm Prim}}{\rmd |t|}\,,
\end{split}
\ee
with a common proportionality factor that cancels in \eqref{eq:lamth}. The result can be conveniently written in terms of the ratio $r$ from \eqref{eq:r}
\be
\lambda_\theta 
= \frac{2r - 1}{2r + 3}\,.
\label{eq:lamthr}
\ee
In the NR limit $\lambda_\theta$ will be completely independent of $|R'(0)|$ and thus sensitive only to the gluon Sivers function. The two interesting cases of \eqref{eq:lamthr} are $r\to 0$ (Primakoff contribution only with transverse $\chi_{c1}$) and $1/r \to 0$ (Sivers contribution only with longitudinal $\chi_{c1}$), where $\lambda_\theta \to -1/3$ and $\lambda_\theta \to 1$, respectively. Hence, even the sign of $\lambda_\theta$ could indicate the presence of the spin dependent Odderon.

In fig.~\ref{fig:lamth} we show $\lambda_{\theta}(x)$ for the above-mentioned models for the gluon Sivers functions. For the ``SIDIS2" model the Sivers channel is small (see fig.~\ref{fig:primsiv2}) and drops with $x$ ($r\ll 1$) and so $\lambda_\theta$ is numerically almost at the pure Primakoff value of $-1/3$. 
The ``5\% positivity" model shows the opposite behavior, with a slightly negative $\lambda_\theta$ at $x\le 0.1$ which increases towards the pure Sivers value of $+1$ at $x\sim 10^{-3}$.

\section{Discussion and conclusion}
We suggest that the measurement of the exclusive $\chi_{c1}$ production cross section and polarization in the forward
limit of high-energy unpolarized $\gamma- p$ collisions can provide stringent constraints on the gluon Sivers function $f_{1T}^{\perp g}(x,\kp)$ or, alternatively, on the
$h_{1T}^g(x,\kp)$ and $h_{1T}^{\perp g}(x,\kp)$ TMDs which at small $x$ are equal to $f_{1T}^{\perp g}(x,\kp)$
\cite{Boer:2015pni}. In fact, the cross section \eqref{eq:sivch}, being proportional to the {\it square} of the first moment of the Sivers function in the NR limit, is potentially even more sensitive than the traditional SSAs with a {\it linear} dependence on $f_{1T}^{\perp g}(x,\kp)$.
The gluon Sivers function is a key element of our understanding of the proton and, to date, is very poorly known.
Based on the connection between the spin dependent Odderon and $f_{1T}^{\perp g}(x,\kp)$ we compute the cross section and find a non-zero contribution at $t\to 0$. Thanks to the Landau-Yang theorem the background from the Primakoff channel is finite. 
Some current models for the gluon Sivers function suggest that the two channels may have cross sections of similar magnitude. By examining angular distributions of the $\chi_{c1}$ decay products we find that the $\cos^2\theta$ angular coefficient in the Primakoff and Sivers channels, has opposite sign.
In conjunction with similar measurements of $C$-odd $t$-channel exchanges at high momentum transfer~\cite{Benic:2024pqe}
the processes discussed here have great potential to provide fundamental insight into the structure of the proton.

\begin{acknowledgments}
We thank Y.~Hatta for useful comments on the manuscript. S.~B. thanks Z.~Tu for discussion on polarization. The work of S.~B.\ is supported by the Croatian Science Foundation (HRZZ) no. 5332 (UIP-2019-04). A.~D.\ acknowledges support
by the DOE Office of Nuclear Physics through Grant DE-SC0002307. T.~S.~kindly acknowledges the support of the Polish National Science Center (NCN) Grant No.\,2021/43/D/ST2/03375. S.~B. acknowledges the hospitality of the EIC Theory Institute at the Brookhaven National Laboratory in March 2024 where this work was initiated.
\end{acknowledgments}

\appendix

\section{Connection to the tri-gluon correlator at small-$x$}
\label{sec:appa}

Here we collect relations between the spin dependent Odderon, the tri-gluon collinear correlator, and the gluon TMDs. Though these relations are available in the literature \cite{Zhou:2013gsa,Boer:2015pni,Dong:2018wsp} we consider it useful to summarize them here.  For simplicity we work in the transverse spin basis. We start with the gluon forward matrix element \cite{Boer:2015pni,Boer:2016xqr,Boer:2018vdi}\footnote{Our conventions for the definition of this correlator
follow ref.~\cite{Boer:2018vdi}; see their eq.~(3.1) for which we use 
$\Gamma^{ij}(x,\kp) = G^{[+.-]ij}(x,\kp,\xi = 0, \delp = 0)$.}
\be
\Gamma^{ij}(x,\kp) = \frac{2}{P^+}\int \frac{\rmd v^- \rmd^2 \vp}{(2\pi)^3} \rme^{\rmi x P^+ v^-} \rme^{-\rmi \kp\cdot \vp} \langle P \Sp | {\rm tr}\big[F^{i+}(-v/2) \calU^{[+]}_{[-\frac{v}{2},\frac{v}{2}]} F^{j+}(v/2)\calU^{[-]\dagger}_{[-\frac{v}{2},\frac{v}{2}]}\big]|P \Sp\rangle\,,
\label{eq:gltmd}
\ee
where $\calU_{[x,y]}^{[\pm]}$ are gauge link staples $\calU_{[x,y]}^{[\pm]} = [x^-,\pm \infty]_{\xp} [\xp,\yp]_{\pm \infty} [\pm \infty,y^-]_{\yp}$ so that \eqref{eq:gltmd} represents the so-called dipole TMD \cite{Dominguez:2011wm}. The notation $[x^-,\pm \infty]_{\xp}$ stands for
\be
[x^-,\pm \infty]_{\xp} = \calP \exp\left[-\rmi g \int_{x^-}^{\pm \infty} \rmd y^- A^+(y^+ = 0,y^-,\yp = \xp)\right]\,,
\ee
while $[\xp,\yp]_{\pm \infty}$ denote the transverse gauge links. We have the following decomposition \cite{Mulders:2000sh,Meissner:2007rx,Boer:2015pni}
\be
\Gamma^{ij}(x,\kp) = x f_1^g(x,\kp)\frac{1}{2}\delta^{ij} + x f^{\perp g}_{1T}(x,\kp)t_1^{ij} +  x h^g_{1T}(x,\kp)t_2^{ij} + x h_{1T}^{\perp g} (x,\kp)t_3^{ij}+\dots\, ,
\label{eq:gamma}
\ee
where
\be
\begin{split}
& t_1^{ij} = -\frac{1}{2}\delta^{ij}\frac{(\kp\times \Sp)}{M_p}\,,\\
& t_2^{ij} = - \frac{1}{4 M_p} \left(\epsilon^{il}k_\perp^l S_\perp^j + \epsilon^{jl}k_\perp^l S_\perp^i + \epsilon^{il}k_\perp^j S_\perp^l + \epsilon^{jl}k_\perp^i S_\perp^l\right)\,,\\
& t_3^{ij} = \frac{(\kp\cdot \Sp)}{2 M_p\kp^2}\left(\epsilon^{il}k_\perp^l k_\perp^j + \epsilon^{jl}k_\perp^l k_\perp^i\right)\,.
\end{split}
\ee
Terms contained in the ellipses are irrelevant in the following. In addition to the gluon Sivers function $f_{1T}^{\perp g}(x,\kp)$, two additional TMDs: $h^g_{1T}(x,\kp)$ and $h_{1T}^{\perp g}(x,\kp)$ emerge for a full description of the transversely polarized proton at leading twist \cite{Mulders:2000sh,Meissner:2007rx,Boer:2015pni}. Our normalization convention for the TMDs differs from \cite{Boer:2015pni} by a factor of $\frac{1}{2}$, and is in line with ref.~\cite{Yao:2018vcg}, so that the unpolarized gluon PDF is obtained as $f_1^g(x) = \int \rmd^2 \kp f_1^g(x,\kp)$. Following the steps in ref.~\cite{Boer:2018vdi} (for similar derivations, see also \cite{Dominguez:2011wm,Fu:2024sba}), the small-$x$ limit of \eqref{eq:gltmd} becomes
\be
\Gamma^{ij}(x,\kp) = \frac{4 N_c}{(2\pi)^3 g^2}k_\perp^i k_\perp^j \int_{\xp\yp} \rme^{-\rmi \kp(\xp - \yp)} \frac{1}{N_c}\frac{\langle P \Sp |{\rm tr}[V(\xp)V^{\dag}(\yp)] | P\Sp\rangle}{\langle P \Sp | P \Sp\rangle}\,.
\ee
where $V(\xp) = [-\infty,+\infty]_{\xp}$ and we have used that, in covariant gauge, the transverse gauge links can be dropped.

Considering now the combination $\Gamma^{ij}(x,\kp) - \Gamma^{ji}(x,-\kp)$, we get
\be
\begin{split}
\frac{1}{2}\left[\Gamma^{ij}(x,\kp) - \Gamma^{ji}(x,-\kp)\right] & = -\frac{4\rmi N_c}{(2\pi)^3 g^2} k_\perp^i k_\perp^j \int_{\xp\yp} \rme^{-\rmi \kp\cdot(\xp - \yp)} \calO_{\Sp\Sp}(\xp,\yp)\\
& = \frac{2}{\pi g^2}k_\perp^i k_\perp^j\frac{(\kp\times \Sp)}{M_p} O_{1T}^\perp(x,\kp)\,,
\label{eq:oddtmd}
\end{split}
\ee
where $\calO_{\Sp\Sp}(\xp,\yp)$ is the Odderon exchange amplitude which has the same form as in eq.~\eqref{eq:Odef} but for a transversely polarized proton. In the second line we have used the fact that the Fourier transform of $\calO_{\Sp\Sp}(\xp,\yp)$ is parametrized in terms of a unique combination: the spin-dependent Odderon $\propto (\kp\times\Sp) O_{1T}^\perp(x,\kp)$ \cite{Boer:2015pni,Zhou:2013gsa}.

We now review briefly the result of Ref.~\cite{Boer:2015pni} who showed that $f_{1T}^{\perp g}(x,\kp)$, $h^g_{1T}(x,\kp)$ and $h_{1T}^{\perp g}(x,\kp)$ are all equal at small-$x$. We compute $t_{l}^{ij}[\Gamma^{ij}(x,\kp) - \Gamma^{ji}(x,-\kp)]$ from \eqref{eq:oddtmd} and also from \eqref{eq:gamma} and compare the two results. Using the relations
\be
\begin{split}
& k_\perp^i k_\perp^j t_1^{ij} = k_\perp^i k_\perp^j t_2^{ij} = -\frac{1}{2}\kp^2 \frac{(\kp\times \Sp)}{M_p}\,,\\
& k_\perp^i k_\perp^j t_3^{ij} = 0\,,
\end{split}
\ee
and also
\be
\begin{split}
& t_1^{ij}t_1^{ij} = \frac{1}{2 M_p^2}(\kp\times\Sp)^2 = \frac{1}{2 M_p^2}(\kp^2 - (\kp \cdot \Sp)^2)\,,\\ 
& t_1^{ij} t_2^{ij} = 0\,,\\
& t_1^{ij} t_3^{ij} = 0\,,\\
& t_2^{ij} t_2^{ij} = \frac{1}{2 M_p^2} \kp^2\,,\\
& t_2^{ij} t_3^{ij} = -\frac{1}{2 M_p^2}(\kp\cdot\Sp)^2\,,\\
& t_3^{ij} t_3^{ij} = \frac{1}{2 M_p^2}(\kp\cdot\Sp)^2\,,
\end{split}
\label{eq:projec}
\ee
we have
\be
\begin{split}
 -\frac{1}{\pi g^2} \kp^2 \frac{(\kp\times \Sp)^2}{M_p} O_{1T}^\perp(x,\kp) & =  \frac{1}{2 M_p^2} (\kp\times \Sp)^2 x f_{1T}^{\perp g} (x,\kp)\,,\\
 -\frac{1}{\pi g^2} \kp^2 \frac{(\kp\times \Sp)^2}{M_p} O_{1T}^\perp(x,\kp) & = \frac{1}{2 M_p^2} \left(\kp^2 x h_{1T}^g(x,\kp) - (\kp\cdot \Sp)^2 x h_{1T}^{\perp g}(x,\kp)\right)\,,\\
 0 & = \frac{1}{2M_p^2}(\kp\cdot\Sp)^2 \left(-x h_{1T}^g(x,\kp) + x h_{1T}^{\perp g}(x,\kp)\right)\,,
\end{split}
\ee
from which we deduce
\be
x f_{1T}^{\perp g}(x,\kp) = x h_{1T}^g(x,\kp) = x h_{1T}^{\perp g}(x,\kp)\,, 
\ee
completely confirming the result in \cite{Boer:2015pni,Boer:2016xqr}. This immediately implies equality of the first $k_\perp$-moments: 
\be
x f_{1T}^{\perp (1) g}(x) = x h_{1T}^{(1) g}(x) =  x h_{1T}^{\perp (1) g}(x)\,.
\label{eq:moms2}
\ee

We now establish the connection of the TMD moments \eqref{eq:moms2} to the tri-gluon correlator. For this purpose it is simplest to take the $t_1^{ij}$ projection. According to \eqref{eq:projec} this selects only $x f_{1T}^{\perp (1) g}(x)$ and then the task is to check its operator definition. We have
\be
\begin{split}
M_p x f_{1T}^{\perp (1) g}(x) & = \int \rmd^2 \kp t_1^{ij}\frac{1}{2}\left[\Gamma^{ij}(x,\kp) - \Gamma^{ji}(x,-\kp)\right]\\
& = \frac{4\rmi N_c }{(2\pi)^3 g^2} \int \rmd^2 \kp \kp^2 (\kp\times \Sp) \int_{\xp\yp} \rme^{-\rmi \kp\cdot(\xp - \yp)} \calO_{\Sp\Sp}(\xp,\yp)\\
& = \frac{4\rmi N_c}{(2\pi)^3 g^2}\int \rmd^2 \kp \kp^2 (\kp\times \Sp) \frac{-g^3}{24 N_c} d_{abc}\frac{1}{2 P^+ 2\pi \delta(0)}\\
&\times\int_{\rp} \rme^{-\rmi \kp\cdot\rp}\big\langle P \Sp| 3 \alpha_a(\rp/2)\alpha_b(-\rp/2)\alpha_c(-\rp/2) - 3\alpha_a(\rp/2)\alpha_b(\rp/2) \alpha_c(-\rp/2)\\
& + \alpha_a(\rp/2)\alpha_b(\rp/2)\alpha_c(\rp/2) - \alpha_a(-\rp/2)\alpha_b(-\rp/2)\alpha_c(-\rp/2)|P \Sp \rangle\,.
\end{split}
\label{eq:momsiv}
\ee
To get the third equality we have used eq.~\eqref{eq:oddexp} for $O_{\Sp\Sp}(\xp,\yp)$,  and translation invariance to shift the dipole coordinates $\xp = \bp + \rp/2$ and $\yp = \bp - \rp/2$ by $\bp$. We recall here that $\alpha_a(\xp)$ is a shorthand for the gluon field integrated along the light-cone $x^-$ coordinate, see eq.~\eqref{eq:alpa}. We now Fourier transform each $\alpha_a(\pm\rp/2)$ and arrange the result in the following way
\be
\begin{split}
M_p x f_{1T}^{\perp (1) g}(x) & = -\frac{\rmi}{96\pi}\frac{g}{P^+} \frac{1}{2\pi \delta(0)}d_{abc}\int_{\konp\ktwp\kthp}d_{abc}\langle P\Sp|\alpha_a(\konp)\alpha_b(\ktwp)\alpha_c(\kthp)|P\Sp\rangle\\
& \times \big[(\konp\times\Sp)(\konp^2  + \ktwp^2 + \kthp^2 - 4(\konp\cdot\ktwp) + 2 (\konp\cdot \kthp) + 8 (\ktwp\cdot\kthp))\\
& - 2(\ktwp\times\Sp)(\konp^2  + \ktwp^2 + \kthp^2 - (\konp\cdot\ktwp) - 4 (\konp\cdot \kthp) - (\ktwp\cdot\kthp))\\
& +(\kthp\times\Sp)(\konp^2  + \ktwp^2 + \kthp^2 + 8(\konp\cdot\ktwp) +2  (\konp\cdot \kthp) - 4 (\ktwp\cdot\kthp))\big]\,.
\end{split}
\label{eq:oddmom}
\ee
We can easily confirm that the combination $(\konp\times \Sp) \konp^2 - 2 (\ktwp\times \Sp)\ktwp^2 + (\kthp\times \Sp)\kthp^2$ sums to zero by relabeling the integration variables as $\ktwp \to \konp$ and $\kthp \to \konp$ in the second and the third term, and by using the symmetry of the $d_{abc}$-factor. In fact, by a similar analysis, all the terms in the square bracket in \eqref{eq:oddmom} cancel out, except the terms with the structure $(\konp\times \Sp)(\ktwp\cdot\kthp)$ and its permutations. In total we end up with
\be
M_p x f_{1T}^{\perp (1) g}(x) = -\frac{\rmi}{4\pi}\frac{g}{P^+} \frac{1}{2\pi \delta(0)}\int_{\konp\ktwp\kthp} (\konp\times\Sp)(\ktwp\cdot\kthp) d_{abc} \langle P \Sp|\alpha_a(\konp) \alpha_b(\ktwp)\alpha_c(\kthp)|P \Sp\rangle\,.
\ee
Now we perform the momentum integrals to obtain the collinear field strength as
\be
\int_{\kp} k_{\perp}^i \alpha_a(\kp) = \rmi \int \rmd x^- F_a^{i+}(x^-)\,,
\ee
where $F^{i+}_a = \partial^i A_a^+$. Using $2\pi \delta(0) = \int^\infty_{-\infty} \rmd x^-$, and again employing translation invariance, we find
\be
\begin{split}
M_p x f_{1T}^{\perp (1) g}(x) & = - \frac{1}{4\pi} \epsilon^{ik}S_\perp^k \delta^{jl}\frac{g}{P^+}\int_{-\infty}^\infty \rmd y^- \int_{-\infty}^\infty\rmd z^- d_{abc}\langle P \Sp|F_a^{i+}(0)F_a^{l+}(y^-)F_a^{j+}(z^-)| P \Sp\rangle\\
& = 2\pi M_p O(x)\epsilon^{ik} S_\perp^k\delta^{jl} S_\perp^r(\delta^{ij} \epsilon^{lr} + \delta^{il} \epsilon^{jr} + \delta^{jl}\epsilon^{ir})\,,
\end{split}
\ee
where we used the definition \eqref{eq:trig} of the tri-gluon correlator.
Performing the contractions we obtain the final result
\be
x f_{1T}^{\perp (1) g} = 8\pi O(x)\,.
\ee

\typeout{}
\bibliography{references}

\end{document}